\documentclass[conference]{IEEEtran}
\IEEEoverridecommandlockouts
\usepackage{cite}
\usepackage{amsmath,amssymb,amsfonts,amsthm}
\usepackage{algorithmic}
\usepackage{graphicx}
\usepackage{textcomp}
\usepackage{xcolor}

\newcommand{\lD}{\mathcal{D}_{\mathrm{low}}}
\newcommand{\miD}{\mathcal{D}_{\mathrm{mid}}}
\newcommand{\hD}{\mathcal{D}_{\mathrm{high}}}

\usepackage{amsthm}
\usepackage{xcolor}
\usepackage{mathtools}
\usepackage{pifont}
\usepackage[OT2,T1]{fontenc}

\usepackage{thmtools}
\usepackage{thm-restate}
\usepackage{paralist}

\newtheorem{theorem}{Theorem}[section]

\newtheorem{definition}{Definition}[section]
\newtheorem{lemma}[theorem]{Lemma}
\newtheorem{corollary}{Corollary}[section]

\DeclarePairedDelimiterX{\infdivx}[2]{[}{]}{%
  #1\;\delimsize\|\;#2%
}
\DeclarePairedDelimiterX{\inftvd}[2]{(}{)}{%
  #1\;\delimsize\|\;#2%
}
\DeclarePairedDelimiterX{\infdivcolon}[2]{[}{]}{%
  #1\;\delimsize;\;#2%
}
\DeclarePairedDelimiterX{\infdivent}[1]{[}{]}{%
  #1
}

\DeclarePairedDelimiterX{\innerProd}[2]{\langle}{\rangle}{%
    #1,#2%
}

\newcommand{\Var}{\mathrm{Var}}

\newcommand{\Ind}{\mathbf{1}}

\DeclareSymbolFont{cyrletters}{OT2}{wncyr}{m}{n}
\DeclareMathSymbol{\Sha}{\mathalpha}{cyrletters}{"58}
\DeclareMathSymbol{\sha}{\mathalpha}{cyrletters}{"78}

\usepackage{amsmath,amsfonts,bm}

\def\eqref#1{equation~\ref{#1}}
\def\1{\bm{1}}

\DeclareMathAlphabet{\mathsfit}{\encodingdefault}{\sfdefault}{m}{sl}
\SetMathAlphabet{\mathsfit}{bold}{\encodingdefault}{\sfdefault}{bx}{n}

\newcommand{\R}{\mathbb{R}}

\DeclareMathOperator*{\esup}{ess\,sup}

\newenvironment{restatedlemma}[1]{%
  \par\medskip
  \noindent\textbf{Lemma~\ref{#1}}\itshape
}{%
  \par\medskip
}

\newcommand{\BallSmallest}{
  Given a realization $(y^n,u)$, let $A \subset \mathcal{X}^n$ be an arbitrary nonempty set and let $B \in \mathcal{X}^n$ be a nonempty distortion ball s.t.
    \begin{align}
        E_{X^n|A}\left[ d(X^n,y^n)|A\right] \le E_{X^n|B}\left[ d(X^n,y^n)|B\right].
    \end{align}
    Then,
    \begin{equation}
        P_{X^n}(A) \le P_{X^n}(B).
    \end{equation}
}

\newcommand{\SymmetricIsCurved}{
  For a symmetric source-distortion pair, the rate-distortion function $R_I(D)$ is curved about any $0<D<D^*$.
}

\newcommand{\StrictConResult}{
  Given that the rate-distortion function is curved at $D$, consider a distortion level $\hat{D} \in (0, D^*] \setminus [D_1, D_2]$. Then, we have
\begin{align}
    \Delta := R_I(\hat{D}) - R_I(D) - R_I'(D)\left( \hat{D} - D\right) \ge H.
\end{align}
}
\def\BibTeX{{\rm B\kern-.05em{\sc i\kern-.025em b}\kern-.08em
    T\kern-.1667em\lower.7ex\hbox{E}\kern-.125emX}}
\begin{document}

\title{Exact Redundancy for Symmetric Rate-Distortion

\thanks{This research was supported by the US National Science 
    Foundation under Grant CCF-2306278.
}
}

\author{\IEEEauthorblockN{Sharang M. Sriramu}
\IEEEauthorblockA{\textit{School of Electrical and Computer Engineering} \\
\textit{Cornell University}\\
Ithaca, USA \\
sms579@cornell.edu}
\and
\IEEEauthorblockN{Aaron B. Wagner}
\IEEEauthorblockA{\textit{School of Electrical and Computer Engineering} \\
\textit{Cornell University}\\
Ithaca, USA \\
wagner@cornell.edu}
}

\maketitle
\begin{abstract}
For variable-length coding with an almost-sure distortion
constraint, Zhang \emph{et al.}~\cite{zhang1997redundancy} show that for discrete sources the redundancy is upper bounded by $\log n/n$ and lower
bounded (in most cases) by $\log n/(2n)$, ignoring lower order
terms. For a uniform source with a distortion measure satisfying
certain symmetry conditions, we show that $\log n/(2n)$ is achievable and that this cannot be improved even if one relaxes the distortion constraint to be in expectation rather than with probability one.
\end{abstract}

\section{Introduction}
Classical rate-distortion theory \cite{berger2003rate} characterizes the smallest achievable compression rate for a stochastic source, subject to the reconstruction satisfying a distortion constraint. The canonical problem in this framework is the compression of a discrete memoryless source under a single-letter distortion criterion. In the setting of a variable-length, prefix-free code, which is the focus of this paper, this problem can be formulated with either an \emph{average-distortion} constraint or the more restrictive \emph{$d$-semifaithful}~~\cite{yu1993rate},~\cite{ornstein1990universal},~\cite{10061340},~\cite{9646872} constraint. The former requires that the expected block distortion not exceed a prescribed threshold, while the latter requires the block distortion constraint to be satisfied with probability one.

It is a standard result that the rate-distortion function is
the asymptotically optimal rate and is a lower bound on the
rate at any blocklength, under both constraints~\cite{ThomasCover,csiszar2011information}. More recent work has focused on characterizing the \emph{redundancy}, which is defined as the gap between the optimal rate and the rate-distortion function as
a function of the blocklength.

In the $d$-semifaithful setting, Zhang \emph{et al.} \cite{zhang1997redundancy} show that the redundancy is bounded between $\frac{\log n}{2n}$ \footnote{Throughout this paper, $\log(\cdot)$ refers to the natural logarithm, and $\exp(x)$ denotes $e$ raised to the power of $x$.} and $\frac{\log n}{n}$, ignoring
lower-order terms here and throughout this discussion. Their achievability proof uses a random coding argument where the encoder selects the first codeword from a random codebook that satisfies the distortion constraint then communicates its index losslessly using the Elias code \cite{elias2003universal}. The converse relies on a combinatorial analysis of type-dependent distortion balls, yielding tight bounds on the size of the pre-image sets induced by any admissible code. This part of the analysis relies on the uniqueness of the optimal output distribution. Earlier,
Merhav~\cite{391272} showed a $\frac{\log n}{2n}$ lower bound
for the case in which the distortion measure is Hamming 
distance. Thus the redundancy for $d$-semifaithful coding is
open although tightly bounded.
For the expected distortion constraint, it appears
that no redundancy bounds are known other than the trivial
lower bound of zero and the upper bound of $\frac{\log n}{n}$
that comes from the $d$-semifaithful case. It should be noted,
however, that the redundancy is known exactly for the fixed-rate,
expected distortion case~\cite{75241}. Results are also known
in various settings for which the redundancy is super-logarithmic,
such as minimax universal $d$-semifaithful coding~\cite{10299682} and 
coding with an excess distortion probability constraint~\cite{6145679}.

Returning to the setup under study,
we focus on symmetric source–distortion pairs in which the source distribution is uniform and the distortion criterion is permutation symmetric (see Definition~\ref{def:SymmetricRDPair} to follow). For this class, we 
determine the redundancy in a strong sense: we show that 
$\frac{\log n}{2n}$ is achievable with $d$-semifaithful coding
and that this cannot be improved even if we relax to an
expected distortion constraint. We thus show that the 
Zhang \emph{et al.} lower bound for $d$-semifaithful coding
cannot be improved in general. We also provide the first nontrivial redundancy lower bound
for a class of sources under an expected distortion constraint (with variable-length coding).
%In this regime, we show that a redundancy of $\frac{\log n}{2n}$ is both achievable and optimal under the $d$-semifaithful constraint, thereby closing the gap in \cite{zhang1997redundancy}. We further extend the converse to the average-distortion setting, showing that the redundancy cannot be reduced even when the distortion constraint is relaxed.

Our achievability proof hinges on the observation that, under symmetry, the probability of a distortion ball depends only on its radius and is invariant to its center. The distribution of the selected index in Zhang \emph{et al.}'s \cite{zhang1997redundancy} random coding argument is independent of the source sequence, and hence becomes known to the decoder. This decoupling eliminates the need for a universal integer code: the index can instead be conveyed using an entropy code matched to its known distribution, yielding a strict redundancy improvement of $\frac{\log n}{2 n}$.

Another way to understand this is from the perspective of \emph{channel simulation}~\cite{li2024channel}. The random coding scheme \emph{simulates} a channel which maps a source sequence to a uniformly random reconstruction within its target distortion ball. Under symmetry, this induced channel is \emph{singular}~\cite{sriramu2024optimal},~\cite{altuug2020exact}, which is known to admit low-redundancy simulation. The redundancy savings above are thus a direct manifestation of singularity in the simulated channel.

For the converse, we use a proof strategy similar to the channel simulation converse proof in Flamich \emph{et al.} \cite{flamich2025redundancy} to replace the pre-image sets of an admissible code by commensurate distortion balls, which can then be characterized using large-deviations techniques. The converse
argument also relies on the fact that the probability of
a distortion ball is invariant to its center.

Since our assumptions on the source and distortion measure
are somewhat restrictive, it is worth noting that our lower
bound does not hold for arbitrary source and distortion
pairs. Specifically, it does not hold if
the rate–distortion function has a linear segment at the zero-rate endpoint, as shown in section~\ref{sec:straight}.
We show as part of our proof (see Lemma~\ref{lemma:SymmetricIsCurved}) that a symmetric source–distortion pair induces a \emph{curved} rate–distortion function, ruling out linear segments near the boundary points.

The full version of this paper, which contains an appendix with the proofs and calculations omitted here is available on the arXiv \cite{arxiv_version}.

\section{Problem Setup} \label{section:ProblemSetupRD}
Consider two finite sets $\mathcal{X}$ and $\mathcal{Y}$ representing the source and reconstruction alphabets, respectively, with $|\mathcal{X}| = |\mathcal{Y}|$. Let $X \sim P_X$, taking values in $\mathcal{X}$, represent the \emph{source} and the mapping $d: \mathcal{X} \times \mathcal{Y} \mapsto [0, D_{\text{max}})$ represent the distortion metric. We extend our distortion metric to operate on block length $n$ sequences by defining $d(x^n, y^n) = \frac{1}{n}\sum\limits_{i=1}^n d(x_i, y_i)$.

\begin{definition} \label{def:SymmetricRDPair}
    A source-distortion pair $(P_X, d)$ is said to be \emph{symmetric} if it satisfies the following:
\begin{enumerate}[i)]
    \item The matrix corresponding to $d(\cdot, \cdot)$ is symmetric, i.e., each row is a permutation of the other rows, and each column is a permutation of the other columns. \label{assumption:distSymmRD}
    \item The source distribution $P_X$ is uniform over $\mathcal{X}$.
    \item For any $y \in \mathcal{Y}$, the distortion $d(\cdot, y)$ is not identically $0$. This ensures that $\sigma^2 := \min\limits_{y \in \mathcal{Y}} \Var_{P_X} \left[ d(X, y)\right] >0$, \label{assumption:UniformSourceRD}
    \item $\max\limits_{x \in \mathcal{X}}\min\limits_{y \in \mathcal{Y}}d(x, y) = 0$, \text{ and}
\end{enumerate}
\end{definition}

\begin{definition}
    The \emph{common randomness} is denoted by a random variable $U$ with distribution $P_U$, taking values in some set $\mathcal{U}$, and independent of the source $X$. A \emph{stochastic scheme} consists of the triplet $(f, g, U)$ defined by the mappings $f:\mathcal{X}^n \times \mathcal{U} \mapsto \{0,1\}^*$ and $g: \{0,1\}^* \times \mathcal{U} \mapsto \mathcal{Y}^n$. The encoder $f$ is additionally restricted to output prefix-free strings.
    \begin{enumerate}[i)]
        \item The scheme achieves an \emph{average} rate-distortion performance of $(R,D)$ if 
        \begin{align}
            E_{X^n,U}\left[ d(X^n, g(f(X^n, U), U)) \right] &\le D, \text{ and}\\
            E_{X^n,U}\left[ \frac{1}{n}\ell\left( f(X^n, U) \right) \right] &\le R.
        \end{align} 
        We will define the operational rate-distortion function for the stochastic average distortion scheme as
        \begin{align}
            R_n(D) = &\inf\limits_{f,g,U} E_{X^n,U}\left[ \frac{1}{n}\ell\left( f(X^n,U) \right) \right] \nonumber\\
            &s.t. \nonumber\\
            &E_{X^n,U}\left[ d(X^n, g(f(X^n,U),U)) \right] \le D.
        \end{align}
        \item The scheme achieves a $d$-semifaithful rate-distortion performance of $(R,D)$ if
        \begin{align}
            d(X^n, g(f(X^n, U), U)) &\le D, \, P_{X}^{\times n} \times P_U \text{ a.s., and}\\
            E_{X^n,U}\left[ \frac{1}{n}\ell\left( f(X^n, U) \right) \right] &\le R.
        \end{align}
        The $d$-semifaithful operational rate-distortion function is defined as
        \begin{align}
            \overline{R_n}(D) = &\inf\limits_{f,g,U} E_{X^n,U}\left[ \frac{1}{n}\ell\left( f(X^n,U) \right) \right] \nonumber\\
            &s.t. \nonumber\\
            &d(X^n, g(f(X^n,U),U)) \le D\,\text{ a.s.}
        \end{align}
    \end{enumerate}
    Note that for all $D$, $\overline{R_n}(D) \ge R_n(D)$.
\end{definition}
Stochastic schemes subsume deterministic schemes, where $U$ is forced to be deterministic, for both the above problems. 
\subsection{Rate-Distortion function} \label{section:Assumptions}
\begin{definition}
    For all $D>0$, the \emph{rate-distortion} function is given by
    \begin{align}
        R_I(D) =& \min\limits_{P_{Y|X}} I(X;Y) \nonumber\\
        &s.t. \nonumber\\
        &E_XE_{Y|X}\left[ d(X, Y)|X \right] \le D. \label{eq:RDDefn}
    \end{align}
    We will also define the \emph{cutoff distortion} $D^* = E_{X}\left[d(X,y)\right]$ for any $y \in \mathcal{Y}$. We observe that $D<D^*$ implies $R(D)>0$. Additionally, we will denote the derivative of the rate-distortion function by $R_I'(\cdot)$.
\end{definition}
We will also find it useful to define a particular notion of \emph{curvature} for the rate-distortion function.

\begin{definition}[Curved rate-distortion function] \label{def:CurvedRD}
    The rate-distortion function is \emph{curved} about a distortion level $0 < D < D^*$ if there exists $0 < D_1 < D$ and $D < D_2 < D^*$ s.t.
    \begin{align}
        R_I(D_i) > R_I(D) + R_I'(D)\left( D_i - D\right) \text{ for } i \in \{1,2\}. \label{eq:CurvatureEqRD}
    \end{align}
\end{definition}

%\section{Previous Results}
%Zhang \emph{et al.} \cite{567651} provided the following bounds on the second-order redundancy for finite alphabet sources\footnote{Here, the limit is to be interpreted in an extended sense, namely via the liminf and limsup.}:
%\begin{align}
%    \frac{1}{2} \le \lim\limits_{n \to \infty} \frac{\overline{R_n}(D) - R_I(D)}{\frac{\log n}{n}} &\le 1,
%\end{align}
%with the lower bound requiring that the optimal output distribution is unique.
%
%Their results hold in a more general setting than we consider here, as they do not assume that the distortion metric is symmetric and that the source distribution is uniform.

\section{Main Result}
We partially address the logarithmic redundancy gap in Zhang \emph{et al.} \cite{zhang1997redundancy} in our work by providing a tighter achievability bound under the additional assumption of symmetric source-distortion pairs (see Definition~\ref{def:SymmetricRDPair}). We also improve the converse under these assumptions by showing that the $\frac{\log n}{n}$ redundancy is tight even if an average distortion constraint is imposed instead of the more restrictive $d$-semifaithful constraint.
\begin{theorem}
    For a symmetric source-distortion pair and distortion level $D$ s.t. $0 < D < D^*$, we have
    \begin{align}
        \lim\limits_{n \to \infty} \frac{\overline{R_n}(D) - R_I(D)}{\frac{\log n}{n}} &= \frac{1}{2}
    \end{align}
    and 
    \begin{align}
        \lim\limits_{n \to \infty} \frac{R_n(D) - R_I(D)}{\frac{\log n}{n}} &= \frac{1}{2}.
    \end{align}
\end{theorem}

\section{Proof of the achievability result}
Given a source sequence $x^n \in \mathcal{X}^n$ observed at the encoder, our approach will be to simulate the following channel 
\begin{align}
    W_{Y^n|X^n}(\cdot|x^n) \propto \Ind\{d(x^n, y^n) \le D\},
\end{align}
which selects an output $y^n$ uniformly at random from the distortion ball with diameter $D$ centered at $x^n$. We will use rejection sampling (see \cite[Section~VI.A]{sriramu2024optimal}) with a uniform proposal distribution $P_Y^{\times n}$ to simulate $W_{Y^n|X^n}$.

Using the common randomness, generate an infinite codebook of realizations drawn i.i.d. from $P_{Y}^{\times n}$:
\begin{equation}
    \mathcal{C}_{\mathrm{RD}} = \{y_1^n, y_2^n,\ldots\}.
\end{equation}
At the encoder, draw i.i.d. samples $U_1, U_2, \ldots$ from $\mathrm{Unif}(0,1)$, select the smallest index $I$ that satisfies
\begin{align}
    U_I &\leq \frac{1}{M_{\mathrm{RD}}}\frac{W_{Y^n|X^n}(y_I^n|x^n)}{P_Y^{\times n}(y_I^n)}, \text{ where}\\
    M_{\mathrm{RD}} &:= \sup\limits_{y^n} \frac{W_{Y^n|X^n}(y^n|x^n)}{P_{Y}^{\times n}(y^n)} = \frac{1}{P_{Y}^{\times n}\left( d(x^n, Y^n) \le D\right)}, \label{eq:M_RD_UB}
\end{align}
and losslessly communicate it to the decoder using $\ell(I)$ bits.

 The symmetry of $d(\cdot,\cdot)$ ensures that all distortion balls having the same diameter have the same cardinality. Therefore, the distortion ball probability $P_{Y}^{\times n}\left( d(x^n, Y^n) \le D\right)$ does not depend on $x^n$. For any $x^n$, we can use the result from~\cite[Lemma~\ref{lemma:ldMain}]{arxiv_version} with $Z_i = -d(x_i, Y_i)$ and $c = -D$ to obtain
 \begin{align}
     P_Y^{\times n}\left( d(x^n, Y^n) \le D\right) &\ge \frac{\underline{M}(\lambda^*, D_{\text{max}}, \sigma^2)}{\sqrt{n}}\exp\left( -n \Lambda^*(-D)\right), \label{eq:LLDRD}
 \end{align}
where the tilting parameter $\lambda^*$ satisfies 
\begin{align}
    \sum\limits_{y^n} \frac{d(x^n, y^n)\exp\left( -\lambda^* d(x^n, y^n)\right)}{\sum\limits_{\hat{y}^n}\exp\left( -\lambda^* d(x^n, \hat{y}^n)\right) } = D
\end{align}
and we have
\begin{align}
    \Lambda^*(-D) &= -\lambda^*D - \Lambda(\lambda^*), \text{ with}\\
    \Lambda(\lambda^*) &= \frac{1}{n}\sum\limits_{i=1}^n \log E_{P_Y}\left[\exp\left( -\lambda^* d(x_i, Y)\right)\right].
\end{align}
By the permutation invariance of $d(\cdot, \cdot)$, this implies
\begin{align}
    \sum\limits_{y} \frac{d(x, y)\exp\left( -\lambda^* d(x, y)\right)}{\sum\limits_{\hat{y}}\exp\left( -\lambda^* d(x, \hat{y})\right) } = D \text{ for all } x.
\end{align}
Define the test channel
\begin{align}
    W^*_{Y|X}(y|x) = \frac{\exp\left( -\lambda^* d(x, y)\right)}{\sum\limits_{\hat{y}}\exp\left( -\lambda^* d(x, \hat{y})\right) }.
\end{align}
We know that $W^*_{Y|X}$ is an optimal test channel (see \cite[Exercise 8.3]{csiszar2011information}) for the rate-distortion problem in (\ref{eq:RDDefn}). From the symmetry of $d(\cdot, \cdot)$, the induced marginal $P_Y(\cdot)$ is uniform over $\mathcal{Y}$.

We can then show that
\begin{align}
    \Lambda^*(-D) = R_I(D). \label{eq:RDOptimalBA}
\end{align}
(The full computation is given in~\cite[Appendix~\ref{section:Detailseq:RDOptimalBA}]{arxiv_version}.)

Combining the bounds in (\ref{eq:LLDRD}) and (\ref{eq:M_RD_UB}) with (\ref{eq:RDOptimalBA}), we obtain
\begin{align}
    M_{\mathrm{RD}} \le \frac{\sqrt{n}}{\underline{M}(\lambda^*, D_{\mathrm{max}}, \sigma^2)} \exp\left( n R_I(D)\right) \label{eq:M_RD_UB2}.
\end{align}
Since we know that the index is geometrically distributed for a fixed source sequence (see Lemma 10, \cite{sriramu2024optimal}) the average amortized length of the optimal code for communicating it can be bounded in terms of $M_{\mathrm{RD}}$, completing the proof:
\begin{align}
    \frac{1}{n} E\left[ \ell(I)\right] &\le \frac{1}{n} \left( \log M_{\mathrm{RD}} + \log(e) + 1\right) \\
    &\le R_I(D) + \frac{\log n}{2 n} + O\left(\frac{1}{n} \right). \label{eq:AchFinalBoundRD}
\end{align}
The details of this computation are described in \cite[Appendix~\ref{section:Detailseq:AchFinalBoundRD}]{arxiv_version}.

\section{Proof of the converse result}
Let $(f, g, U)$ be a stochastic scheme that achieves a rate $R$ and expected distortion $D$. Let us denote the joint distribution induced by this scheme on $\mathcal{X}^n \times \mathcal{Y}^n \times \mathcal{U}$ by $P$.

Using standard information-theoretic inequalities, we obtain
\begin{align}
    nR &\ge H(f(X^n, U)|U)\\
    &\ge I(X^n, f(X^n, U)|U)\\
    &\ge I(X^n;Y^n|U)\\
    &= E_{X^n,Y^n,U}\left[ \log\frac{dP_{X^n|Y^n,U}}{dP_{X^n}} \right]. \label{eq:MappingReduce}
\end{align}
Now, consider the pre-image sets $A(y^n,u) = \{x^n \in \mathcal{X}^n: g(f(x^n,u),u) = y^n\}$. It suffices to examine $(y^n,u)$ realizations where $P_{X^n}(A(y^n,u)) > 0$. For those, the likelihood ratio can be expressed in terms of the pre-image coincidence probability:
\begin{equation}
    \frac{dP_{X^n|Y^n,U}}{dP_{X^n}}(x^n|y^n,u) = \frac{\mathbf{1}\left(x^n \in A(y^n,u)\right)}{P_{X^n}(A(y^n,u))}.
\end{equation}
Substituting this back in (\ref{eq:MappingReduce}) yields
\begin{align}
    nR \ge E_{Y^n,U}\left[ -\log P_{X^n}(A(Y^n, U ) \right]. \label{eq:PreImageBasedBoundRD}
\end{align}
The main thrust of the proof will rely on replacing the pre-image sets $A(y^n,u)$ with \emph{distortion balls}, i.e., sets of the form $\{x^n \in \mathcal{X}^n: d(x^n,y^n) \le t\}$ for some distortion level $t$ --- whose probability can then be characterized using large-deviations results. 

\begin{lemma} \label{lemma:BallSmallest}
    \BallSmallest
\end{lemma}
\begin{proof}
    Please refer to~\cite[{Appendix~\ref{section:LemmaProofs}}]{arxiv_version}.
\end{proof}
Let us define $\tilde{B}(y^n, t)$ to be the smallest distortion ball centered at $y^n$ for which the averaged distortion, \linebreak $E_{X^n|\tilde{B}(y^n,t)}\left[ d(X^n,y^n) | \tilde{B}(y^n, t)\right]$ is greater than or equal to $t$. Let us also define $D(y^n,u) = E_{X^n|A(y^n,u)}\left[ d(X^n, y^n)|A(y^n,u)\right]$. 

Then, applying Lemma~\ref{lemma:BallSmallest} to the bound from (\ref{eq:PreImageBasedBoundRD}), we obtain
\begin{align}
    nR &\ge E_{Y^n,U}\left[ -\log P_{X^n}(\tilde{B}(Y^n, D(Y^n,U) ) \right].
\end{align}
From the column-symmetry of $d(\cdot, \cdot)$ we see that for a fixed $t$, $P_{X^n}(\tilde{B}(y^n, t)$ is invariant w.r.t. $y^n$. Therefore, we have
\begin{align}
    nR &\ge E_{\tilde{D}}\left[ - \log P_{X^n}(\tilde{B}(0^n, \tilde{D} ) \right],
\end{align}
where $0 \in \mathcal{Y}$ is an arbitrary reconstruction symbol and $D(Y^n,U)$ is abbreviated as $\tilde{D}$. We will find it useful to split the range of $\tilde{D}$ into different intervals. For an arbitrary $\epsilon \in \left( 0, \min\left( D_1, D^* - D_2, \frac{H}{1 + |R_I'(D)|} \right) \right)$, we will consider the intervals $\lD = \{\tilde{D} \le \epsilon\}$, $\miD = \{\tilde{D} \in [\epsilon, D^* - \epsilon]\}$, and $\hD = \{\tilde{D} > D^* - \epsilon\}$. Then, using the law of total expectation, we obtain
\begin{align}
    nR &\ge E_{\tilde{D}}\left[ - \log P_{X^n}(\tilde{B}(0^n, \tilde{D} ) \right]\\
    &\ge \Pr\left( \lD \right)\left( - \log P_{X^n}(\tilde{B}(0^n, \epsilon) \right) \nonumber\\
    &\phantom{=} + \Pr\left(\miD\right)E_{\tilde{D}|\miD}\left[  - \log P_{X^n}(\tilde{B}(0^n, \tilde{D} )|\miD \right]. \label{eq:tildeBBound}
\end{align}
The next stage of the proof is to bound the radii of the distortion balls $\tilde{B}(0^n, s)$, for all $s \in \miD$, using a large-deviations argument.

\begin{definition}[Exponentially tilted source distributions]
    Given a tilting parameter $\lambda > 0$ and a reconstruction realization $y \in \mathcal{Y}$, the $\lambda$-tilted source distribution w.r.t. $y$, $Q_{X|y}^{(\lambda)}$, is defined by
    \begin{align}
        \frac{dQ_{X|y}^{(\lambda)}}{dP_X}(x) = \exp\left( - \lambda d(x,y) - \Lambda(\lambda)\right), \text{ where}
    \end{align}
    the $\Lambda(\lambda) = \log E_{X}\left[ \exp(-\lambda d(X, y))\right]$ is the cumulant generating function for an arbitrary $y \in \mathcal{Y}$. For $n$ simultaneous realizations $y^n \in \mathcal{Y}^n$, we extend our definition to $Q_{X^n|y^n}^{(\lambda)} = \prod\limits_{i=1}^n Q_{X|y_i}^{(\lambda)}$. 
    % We will use the following subscript notation to denote expectations under the tilted measure:
    % \begin{align}
    %     E_{Q_{X|y}^{(\lambda)}}\left[\cdot\right] &= E_{X|y,\lambda}\left[\cdot\right], \text{ and}\\
    %     E_{Q_{X^n|y^n}^{(\lambda)}} &= E_{X^n|y^n,\lambda}\left[\cdot\right]. 
    % \end{align}
    
    This is equivalent to the definitions in~\cite[Appendix~\ref{section:ldsetup}]{arxiv_version}, with $Z_i = -d(X_i,y_i)$. We also define the large-deviations rate function (similarly defined in~\cite[Appendix~\ref{section:ldsetup}]{arxiv_version}.):
    \begin{align}
        \Lambda_n^*(b) &= \sup\limits_{\lambda} \left( -\lambda b - \Lambda(\lambda) \right) \text{ for all } b>0.
    \end{align}
\end{definition}

\begin{definition}[Tilting parameter thresholds]
    From the convexity of the cumulant generating function $\Lambda(\cdot)$, for an arbitrary $y \in \mathcal{Y}$, we know that $\Lambda'(\cdot) = -E_{Q_{X|y}^{(\lambda)}}[d(X,y)]$ is monotonically increasing for $\lambda \ge 0$. Using~\cite[Lemma~\ref{lemma:LargeTiltingRegime}]{arxiv_version}, we can also conclude that $\lim\limits_{\lambda \to \infty} E_{Q_{X|y}^{(\lambda)}}\left[ d(X,y) \right] = 0$. Hence, we can define
    \begin{align}
        \overline{\eta} &= \inf\left\{ \lambda > 0: E_{Q_{X|y}^{(\lambda)}}\left[d(X,y) \right] \le D^* - \frac{\epsilon}{2}\right\}, \text{ and}\\
        \underline{\eta} &= \inf\left\{ \lambda > 0: E_{Q_{X|y}^{(\lambda)}}\left[d(X,y) \right] \le \frac{\epsilon}{2}\right\}.
    \end{align}
\end{definition}

\begin{definition}[Replacement radius]
    Define
    \begin{align}
        \imath(s) &= s + \frac{C^*_{\epsilon}}{n}, \text{ where} \label{eq:imath}\\
        C^*_{\epsilon} &= \sup\limits_{\lambda \in [\underline{\eta}, \overline{\eta}]}C^*(\lambda, D_{\text{max}}, \sigma^2). \text{ (see \cite[Lemma~\ref{lemma:GibbsBounded}]{arxiv_version})}
    \end{align}
\end{definition}

Now, consider the distortion ball with radius $\imath(s)$ centered at $y^n$: $B(y^n, \imath(s)) = \left\{ x^n \in \mathcal{X}^n: d(x^n, y^n) \le \imath(s)\right\}$. For \mbox{sufficiently} large $n$, we can apply~\cite[Lemma~\ref{lemma:GibbsBounded}]{arxiv_version} to yield
\begin{align}
    E_{X^n|B(y^n,\imath(s))}\left[ d(X^n, y^n)|B(y^n,\imath(s)) \right] \ge s.
\end{align}
We can subsequently use Lemma~\ref{lemma:BallSmallest} to obtain the bound
\begin{align}
    P_{X^n}(\tilde{B}(y^n, s)) \le P_{X^n}(B(y^n, \imath(s))).
\end{align}
Substituting this in our rate lower bound from (\ref{eq:tildeBBound}), we then have
\begin{align}
    nR &\ge \Pr\left( \lD \right) \left( - \log P_{X^n}(B(y^n, \imath(\epsilon) ) \right) \nonumber\\
    &\phantom{=} + \Pr\left(\miD\right)E_{\tilde{D}|\miD}\left[  - \log P_{X^n}(B(y^n, \imath(\tilde{D}))|\miD \right].
\end{align}
We can now use~\cite[Lemma~\ref{lemma:ldMain}]{arxiv_version} to upper bound these distortion ball probabilities, yielding
\begin{align}
    nR &\ge \Pr\left( \lD \right) \left( \frac{\log n}{2} - \log \overline{M}_\epsilon + n \Lambda_n^*(\imath(\epsilon))\right) \nonumber\\
    & \hspace{-1.5em} + \Pr\left(\miD\right)E_{\tilde{D}|\miD}\left[  \frac{\log n}{2} - \log \overline{M}_\epsilon + n \Lambda_n^*(\imath(\tilde{D}))|\miD \right], \label{eq:BeforeLambda_*}
\end{align}
where $\overline{M}_{\epsilon} = \sup\limits_{\lambda \in [\underline{\eta}, \overline{\eta}]} \overline{M}(\lambda, D_{\text{max}}, \sigma^2).$

In \cite[Appendix~\ref{section:Detailseq:LDRateRDBound}]{arxiv_version}, we show that the large deviations rate function $\Lambda_n^*(s)$ can be lower bounded by the rate-distortion function for all $s \in \miD$: i.e.,
\begin{align}
    \Lambda_n^*(s) \ge R_I(s). \label{eq:LDRateRDBound}
\end{align}
Substituting this in (\ref{eq:BeforeLambda_*}), substituting the shorthand $\imath_{\text{mid}} = E_{\tilde{D}|\miD}\left[\imath(\tilde{D})|\miD\right]$, and applying Jensen's inequality results in
\begin{align}
    nR &\ge \Pr\left( \lD \right) \left( \frac{\log n}{2} - \log \overline{M}_\epsilon + n R_I(\imath(\epsilon))\right) \nonumber\\
    & + \Pr\left(\miD\right) \Big(  \frac{\log n}{2} - \log \overline{M}_\epsilon + n R_I\left( \imath_{\text{mid}} \right) \Big).
\end{align}
\vspace{0.1em}
Observing that $R_I(D^*)=0$, we can add $n \Pr(\hD) R_I(D^*)$ to the lower bound, obtaining
\begin{align}
    nR &\ge \Pr\left( \lD \right) \left( \frac{\log n}{2} - \log \overline{M}_\epsilon + n R_I(\imath(\epsilon))\right) \nonumber\\
    &+ \Pr\left(\miD\right) \left(  \frac{\log n}{2} - \log \overline{M}_\epsilon + n R_I\left(\imath_{\text{mid}}\right) \right) \nonumber\\
    &+ \Pr(\hD)\cdot nR_I(D^*).
\end{align}
Next, we will use the curvature of the rate-distortion function (see Definition~\ref{def:CurvedRD}) to bound its value away from the supporting hyperplane at the target distortion $D$. We will begin by showing that any symmetric source-distortion pair (see Definition \ref{def:SymmetricRDPair}) induces a curved rate-distortion function.

\begin{lemma}\label{lemma:SymmetricIsCurved}
    \SymmetricIsCurved
\end{lemma}
\begin{proof}
   Please refer to~\cite[Appendix~\ref{section:LemmaProofs}]{arxiv_version}.
\end{proof}

Define a curvature gap
\begin{align}
    H := \min\limits_{i \in \{1,2\}} R_I(D_i) - R_I(D) - R_I'(D)\left( D_i - D\right). \label{eq:HDefn}
\end{align}
We can show that the curvature gap outside the interval $[D_1, D_2]$ exceeds $H$.
\begin{lemma} \label{lemma:LocalStrictConvResult}
\StrictConResult
\end{lemma}
\begin{proof}
    Please refer to~\cite[Appendix~\ref{section:LemmaProofs}]{arxiv_version}
\end{proof}
Therefore, dividing by $n$ and using Lemma~\ref{lemma:LocalStrictConvResult} (for $n$ sufficiently large s.t. $\imath(\epsilon) < D_1$), we can obtain
\begin{align}
    R &\ge \Pr\left( \lD \right) \left( R_I(D) + R_I'(D)\left( \imath(\epsilon) - D \right) + H \right) \nonumber\\
    &+ \Pr\left(\miD\right) \left( R_I\left(D\right) + R_I'(D)\left( \imath_{\text{mid}}- D\right) \right) \nonumber\\
    &+ \Pr(\hD)\left(R_I(D) + R_I'(D)(D^* - D) + H\right) \nonumber\\
    &+ \left(1 - \Pr(\hD)\right)\frac{\log n}{2 n} - \frac{\log \overline{M}_\epsilon}{n}.
\end{align}
Using the definition of $\imath(\cdot)$ from (\ref{eq:imath}),
\begin{align}
    R &\ge \Pr\left( \lD \right) \left( R_I(D) + R_I'(D)\left( \epsilon - D \right) + H \right) \nonumber\\
    &+ \Pr\left(\miD\right) \left( R_I\left(D\right) + R_I'(D)\left( E_{\tilde{D}|\miD}\left[\tilde{D}|\miD\right] - D\right) \right) \nonumber\\
    &+ \Pr(\hD)\left(R_I(D) + R_I'(D)(D^* - D) + H\right) \nonumber\\
    &+ \left(1 - \Pr(\hD)\right)\frac{\log n}{2 n} - \frac{\log \overline{M}_\epsilon}{n} + R_I'(D)\frac{C_\epsilon^*}{n}.
\end{align}
We will simplify our notation by denoting $\alpha = \Pr\left( \lD \right)$, $\beta = \Pr\left(\miD\right)$, $\gamma = \Pr\left( \hD \right)$, and $D_{\mathrm{mid}} = E_{\tilde{D}|\miD}\left[\tilde{D}|\miD\right]$. Rearranging terms, we can then write
\begin{align}
    R &\ge R_I(D) + R_I'(D)\left( \alpha \epsilon + \beta D_{\mathrm{mid}} + \gamma D^* - D\right) \nonumber\\
    &+ \alpha H + \gamma H \nonumber\\
    &+ \left(1 - \gamma \right)\frac{\log n}{2 n} - \frac{\log \overline{M}_\epsilon}{n} + R_I'(D)\frac{C_\epsilon^*}{n}. \label{eq:simplifiedForm}
\end{align}
Given that the scheme achieves the distortion bound, we have 
\begin{equation}
    D \ge \beta \cdot D_{\mathrm{mid}} + \gamma \cdot (D^* - \epsilon).
\end{equation}
Substituting this in the bound in (\ref{eq:simplifiedForm}), we obtain
\begin{align}
    R &\ge R_I(D) + \frac{\log n}{2 n} 
      + \alpha\left( H - \epsilon |R_I'(D)| \right) \nonumber\\
    &+ \gamma\left( H - \epsilon|R_I'(D)| - \frac{\log n}{2 n}\right) \nonumber\\
    &- \frac{\log \overline{M}_\epsilon}{n} + R_I'(D)\frac{C_\epsilon^*}{n}.
\end{align}
Choosing $n$ to be sufficiently large such that $\frac{\log n}{2 n} < \epsilon$ and observing that $\epsilon < \frac{H}{1 + |R_I'(D)|}$, we complete the proof.

\section{On Straight Rate-Distortion Functions}
\label{sec:straight}

\begin{figure}
    \begin{center}
    \hspace*{-.2in}
        \scalebox{.5}{\includegraphics{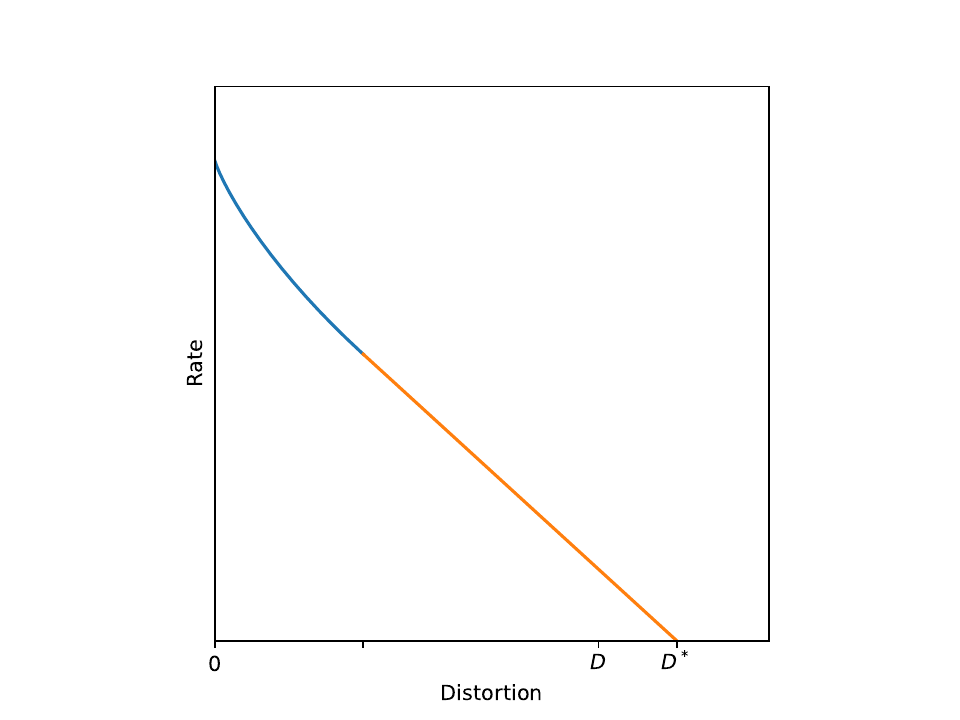}}
    \end{center}
    \caption{A rate-distortion function with a straight-line segment
(orange).}
    \label{fig:RDstraight}
\end{figure}

Suppose the rate-distortion function $R_I(\cdot)$ has a
straight-line segment that ends at $D^*$ as in Fig.~\ref{fig:RDstraight} 
(\hspace{-0.05em}\cite[Exercise~9.4]{Gallager:IT},~\cite{Berger:RD}). Consider $D$ close to $D^*$,
which can be written as
\begin{align}
    R_I(D) & = \theta R_I(D^*) + (1-\theta) R_I(D_0) \\
   & = (1-\theta) R_I(D_0),
\end{align}
where
$\theta D^* + (1-\theta) D_0 = D$
and $0 \le \theta \le 1 $ is close to $1$. Let $y_0$ be
such that $E[d(X,y_0)] = D^*$. Fix $n$, and let $S$ denote
the set of source realizations with probability $\theta$
consisting of those source strings $x^n$ that minimize 
$d(x^n,y_0^n)$, where $y_0^n$ is an $n$-length vector of
$y_0$ symbols. Consider a code that sends one bit to
indicate whether the source realization is in $S$. 
If it is, the decoder outputs $y_0^n$. Otherwise, a
$d$-semifaithful code with distortion $D_0$ and 
rate $R_I(D_0) + \log n/n$ is employed (ignoring lower
order terms). This scheme
meets an expected distortion constraint of $D_0$ with
rate (again ignoring lower-order terms)
\begin{equation}
(1-\theta) \left(R_I(D_0) + \frac{\log n}{n}\right)
 = 
R_I(D) + (1-\theta) \frac{\log n}{n}.
\end{equation}
This redundancy is lower than $\log n/(2n)$ if
$\theta \ge 1/2$. 

\bibliographystyle{IEEEtran}
\bibliography{ref.bib}

\onecolumn
\appendices
\section{Detailed Computations} \label{section:DetailedComputations}

 \subsection{Showing eq. (\ref{eq:RDOptimalBA})} \label{section:Detailseq:RDOptimalBA}
We have
\begin{align}
    &\Lambda^*(-D)\\
    &= -\lambda^*D - \Lambda(\lambda^*)\\
    &= E_{W^*_{Y^n|X^n}}\left[ -\lambda^* d(x^n, Y^n) - \frac{1}{n}\sum\limits_{i=1}^n \log E_{P_Y}\left[\exp\left( -\lambda^* d(x_i, Y)\right)\right] \right]\\
    &= E_{P_{X}}E_{W^*_{Y|X}}\left[ -\lambda^* d(X, Y) - \log E_{P_Y}\left[\exp\left( -\lambda^* d(X, Y)\right)\right] \right]\\
    &= E_{P_{X}}E_{W^*_{Y|X}}\left[ \log \frac{W^*_{Y|X}(Y|X)}{P_Y(Y)}\right] \label{eq:LLRBAForm}\\
    &= R_I(D).
\end{align}
In the above, (\ref{eq:LLRBAForm}) and (\ref{eq:RDOptimalBA}) follows from the definition of the rate-distortion achieving test channel.

 \subsection{Showing eq. (\ref{eq:AchFinalBoundRD})}
\label{section:Detailseq:AchFinalBoundRD}
Using the standard bound on the rate of a prefix-free, lossless code \cite{ThomasCover}, we obtain
\begin{align}
    \frac{1}{n} E\left[ \ell(I)\right] &\le \frac{1}{n} \left( H(I) + 1 \right)\\
    &= \frac{1}{n} \left( H(I|X^n) + 1 \right) \label{eq:NoDepX^n}\\
    &\le \frac{1}{n} \left( \log M_{\mathrm{RD}} + \log(e) + 1\right) \label{eq:GeomEntRD}\\
    &\le \frac{1}{n} \left( \log \left( \frac{\sqrt{n}}{\underline{M}(\lambda^*, D_{\mathrm{max}}, \sigma^2)} \exp\left( n R_I(D)\right) \right) + \log(e) + 1\right) \label{eq:M_RD_UB_follows}\\
    &= R_I(D) + \frac{\log n}{2 n} + O\left(\frac{1}{n} \right).
\end{align}
Here, (\ref{eq:NoDepX^n}) follows from the symmetry of $d(\cdot, \cdot)$, (\ref{eq:GeomEntRD}) follows from Lemma 10 of \cite{sriramu2024optimal}, and (\ref{eq:M_RD_UB_follows}) follows from the bound obtained in (\ref{eq:M_RD_UB2}).

\subsection{Showing eq. (\ref{eq:LDRateRDBound})}
\label{section:Detailseq:LDRateRDBound}

Let $n$ be an integer multiple of $|\mathcal{Y}|$, and $y^n$ be a sequence from the uniform type class, i.e., $\frac{1}{n}\sum\limits_{i=1}^n \mathbf{1}(y_i = \hat{y}) = \frac{1}{|\mathcal{Y}|}$ for all $\hat{y} \in \mathcal{Y}$. Consider the joint measure $Q_{XY}^{(\eta)} = L_{y^n} \times Q_{X|Y}^{(\eta)}$, where $L_{y^n}(\cdot) = \frac{1}{n}\sum\limits_{i=1}^n \mathbf{1}(y_i \in \cdot)$ is the empirical measure w.r.t. $y^n$, and $\eta$ is chosen such that $E_{Q_{X^n|y^n}^{(\eta)}}[d(X^n,y^n)] = s$. The mutual information corresponding to this measure is
\begin{align}
    I(Q_{XY}^{(\eta)}) &= E_{Y \sim L_{y^n}}E_{Q_{X|Y}^{(\eta)}}\left[ \log \frac{dQ_{X|Y}^{(\eta)}}{dQ_{X}^{(\eta)}}\right], \text{where } Q_{X}^{(\eta)}(\cdot) = \frac{1}{n}\sum\limits_{i=1}^nQ_{X|y_i}^{(\eta)}(\cdot).
\end{align}
By symmetry, $Q_{X}^{(\eta)} = P_X$. Then, we have
\begin{align}
    &E_{Y \sim L_{y^n}}E_{Q_{X|Y}^{(\eta)}}\left[ \log \frac{dQ_{X|Y}^{(\eta)}}{dQ_{X}^{(\eta)}}\right]\\
    &= E_{Y \sim L_{y^n}}E_{Q_{X|Y}^{(\eta)}}\left[ \log \frac{dQ_{X|Y}^{(\eta)}}{dP_{X}}\right]\\
    &= \frac{1}{n}\sum\limits_{i=1}^n E_{Q_{X|y_i}^{(\eta)}}\left[ \log \frac{dQ_{X|y_i}^{(\eta)}}{dP_{X}}\right]\\
    &= E_{Q_{X^n|y^n}^{(\eta)}}\left[ - \frac{\eta}{n} \sum\limits_{i=1}^nd(X,y_i) - \Lambda(\eta)\right]\\
    &= -s\eta - \Lambda(\eta)\\
    &= \Lambda_n^*(s).
\end{align}
From this, we can deduce that
\begin{align}
    I(Q_{XY}^{(\eta)}) = \Lambda_n^*(s).
\end{align}
We can then further bound $I(Q_{XY}^{(\eta)})$ from below by the rate-distortion function at distortion level $s$.

\newpage

\section{Proofs of lemmas} \label{section:LemmaProofs}

\begin{restatedlemma}{lemma:BallSmallest}
\BallSmallest
\end{restatedlemma}

\begin{proof} \label{proof:BallSmallest}
    Without loss of generality, we will consider the case $A \not \subset B$ as the result holds trivially otherwise. We will also use the notation $D(A)$ and $D(B)$ to denote $E_{X^n|A}\left[ d(X^n,y^n)|A\right]$ and $E_{X^n|B}\left[ d(X^n,y^n)|B\right]$ respectively. We then have
    \begin{align}
        &D(A) \le D(B)\\
        \implies & D(A \cap B)\frac{P_{X^n}(A \cap B)}{P_{X^n}(A)} + D(A \cap B^C)\frac{P_{X^n}(A \cap B^C)}{P_{X^n}(A)}\\
        \le \,&D(A \cap B)\frac{P_{X^n}(A \cap B)}{P_{X^n}(B)} + D(A^C \cap B)\frac{P_{X^n}(A^C \cap B)}{P_{X^n}(B)}.
    \end{align}
    Because $B$ is a distortion ball, we know that $D(A^C \cap B) \le D(A \cap B^C)$. We therefore have 
    \begin{align}
        \frac{P_{X^n}(A^C \cap B)}{P_{X^n}(B)} &\ge \frac{P_{X^n}(A \cap B^C)}{P_{X^n}(A)} \\
        1 - \frac{P_{X^n}(A \cap B)}{P_{X^n}(B)} &\ge 1 - \frac{P_{X^n}(A \cap B)}{P_{X^n}(A)}\\
        \frac{P_{X^n}(A \cap B)}{P_{X^n}(B)} &\le \frac{P_{X^n}(A \cap B)}{P_{X^n}(A)}\\
        \implies P_{X^n}(A) &\le P_{X^n}(B),
    \end{align}
    completing the proof.
\end{proof}

\begin{restatedlemma}{lemma:SymmetricIsCurved}
\SymmetricIsCurved
\end{restatedlemma}
\begin{proof} \label{proof:SymmetricIsCurved}
    Fix and $0<D<D^*$ and consider any $D_2$ s.t. $D < D_2 < D^*$. We then have,
    \begin{align}
        &R_I'(D) = -\lambda_D, \text{ where } \lambda_D \text{ is the unique } \lambda>0 \text{ s.t.}\\
        &\sum\limits_{y} d(x,y) \frac{\exp\left( -\lambda d(x, y)\right)}{\sum\limits_{y' \in \mathcal{Y}}\exp\left( -\lambda d(x, y')\right)} = D \text{ for all } x\in \mathcal{X}. \label{eq:BARD}
    \end{align}
    Similarly, we have
    \begin{align}
        R_I(D_2) = -\lambda_{D_2}.
    \end{align}
    Now, the left-hand side of (\ref{eq:BARD}) is strictly decreasing in $\lambda$, so
    \begin{align}
        R_I'(D_2) > R_I'(D).
    \end{align}
    We therefore have, for all sufficiently small $\epsilon > 0$,
    \begin{align}
        R_I(D_2 - \epsilon) < R_I(D) - \epsilon R_I'(D). \label{eq:MonotoneUBD2eps}
    \end{align}
    By the convexity of the rate-distortion function, we also have
    \begin{align}
        R_I(D_2 - \epsilon) \ge R_I(D) + R_I'(D)(D_2 - \epsilon - D). \label{eq:ConvexLBD2eps}
    \end{align}
    Combining (\ref{eq:MonotoneUBD2eps}) and (\ref{eq:ConvexLBD2eps}) establishes that
    \begin{align}
        R_I(D_2) > R_I(D) + R_I'(D)(D_2 - D). \label{eq:D2Courv}
    \end{align}
    An analogous argument applies to any $D_1$ s.t. $0<D_1<D$, where we can show that
    \begin{align}
        R_I(D_1) > R_I(D) + R_I'(D)(D_1 - D). \label{eq:D1Curv}
    \end{align}
    Together, (\ref{eq:D2Courv}) and (\ref{eq:D1Curv}) establish the curvature of $R_I(\cdot)$ about $D$.
\end{proof}

\begin{restatedlemma}{lemma:LocalStrictConvResult}
\StrictConResult
\end{restatedlemma}
\begin{proof} 
\label{proof:StrictConResult}
    Let us define
    \begin{align}
    D_\mathrm{th} = \begin{cases} 
        D_1 & \hat{D} < D \\
        D_2 & \hat{D} > D.
   \end{cases}
   \end{align}
    Let $\eta \in (0,1)$ denote the parameter that governs the expression of $D_\mathrm{th}$ as a convex combination of $D$ and $\hat{D}$:
    \begin{align}
        D_\mathrm{th} = (1 - \eta)\cdot D + \eta \cdot \hat{D}. \label{eq:EtaDefn}
    \end{align}
    Then, using the convexity of $R_I(D)$, we obtain
    \begin{align}
        (1 - \eta)\cdot R_I(D) + \eta \cdot R_I(\hat{D}) &\ge R_I(D_\mathrm{th})\\
        \implies (1 - \eta)\cdot R_I(D) + \eta \cdot \left( \Delta + R_I(D) + R_I'(D)\left( \hat{D} - D\right)\right) &\ge R_I(D_\mathrm{th})\\
        \implies \eta \Delta + R_I(D) + \eta\cdot R_I'(D)\left( \hat{D} - D \right) &\ge R_I(D_\mathrm{th})\\
        \implies \Delta &\ge R_I(D_\mathrm{th}) - R_I(D) - \eta\cdot R_I'(D)\left( \hat{D} - D\right) \label{eq:EtaLess1}\\
        \implies \Delta &\ge R_I(D_\mathrm{th}) - R_I(D) - R_I'(D)\left(D_\mathrm{th} - D \right) \label{eq:SubstConvComb}\\
        \implies \Delta &\ge H. \label{eq:D_delta_Defn}
    \end{align}
    Here, (\ref{eq:EtaLess1}) follows because $\eta < 1$ by definition and $\Delta \ge 0$ by convexity of the rate-distortion function. The substitution in (\ref{eq:SubstConvComb}) follows from the identity $\eta \cdot \hat{D} = D_\mathrm{th} - (1 - \eta)\cdot D$ (see the definition in (\ref{eq:EtaDefn})). Finally, (\ref{eq:D_delta_Defn}) follows from (\ref{eq:HDefn})
\end{proof}

\newpage

\section{Large Deviations Results}
\subsection{Exponentially tilted distributions} \label{section:ldsetup}
Given a probability space $(\Sigma, \mathcal{F}, P)$ and positive constants $\sigma^2$ and $\bar{Z}$, consider a random variable $Z:\Omega \to \R$ with $P_Z(\cdot) = P(Z \in \cdot)$ with the following properties
\begin{align}
    |Z| &\le \bar{Z} \text{ P-almost surely, and} \label{eq:stablesource1}\\
    \Var(Z) &\ge \sigma^2. \label{eq:stablesource2}
\end{align}
Given a tilting parameter $\lambda > 0$ and a source $Z$, we can now define the exponentially tilted distribution $P_Z^{(\lambda)}$:
\begin{align}
    \frac{dP_{Z}^{(\lambda)}}{dP_{Z}}(z) = \exp\left(\lambda z - \Lambda(\lambda) \right),
\end{align}
where $\Lambda_Z(\lambda) = \log E_Z\left[ \exp(\lambda Z)\right]$ is the cumulant.\\

We will now state and prove some standard properties of exponentially tilted random variables.

\subsubsection{Properties}
\begin{lemma}[Lower bound on the tilted variance] \label{lemma:TiltedVarianceLowerBound}
For all tilting parameters $\lambda>0$, $\Var_{P_{Z}^{(\lambda)}}[Z] \ge \exp\left( -2\lambda |\bar{Z}| \right)\sigma^2$.
\end{lemma}
\begin{proof}
    Working from the definition of the tilted distribution, we have for all $c$,
    \begin{align}
        \Var_{P_{Z}^{(\lambda)}}[Z] &= E_{P_{Z}^{(\lambda)}}\left[ (Z - c)^2 \right]\\
        &= E_{Z}\left[ \frac{dP_{Z}^{(\lambda)}}{dP_{Z}} \cdot(Z - c)^2 \right]\\
        &= E_{Z}\left[ \exp\left( \lambda Z - \log E_{Z}\left[ \exp(\lambda Z) \right] \right) (Z - c)^2 \right]\\
        &\ge E_{Z}\left[ \exp\left( -2\lambda |\bar{Z}|  \right) (Z - c)^2 \right]\\
        &\ge \exp\left( -2\lambda |\bar{Z}| \right)\sigma^2.
    \end{align}
\end{proof}

\begin{lemma}[Monotonicity of the tilted mean] \label{lemma:TiltedMeanMonotonic}
    The tilted mean,
    \begin{equation}
        \mu(\lambda) = E_{P_{Z}^{(\lambda)}}\left[Z\right],
    \end{equation}
    is strictly monotonically increasing in $\lambda$.
\end{lemma}
\begin{proof}
    We observe that $\mu(\lambda) = \Lambda'(\lambda)$ and that $\Var_{P_{Z}^{(\lambda)}}[Z] = \Lambda''(\lambda)$, which, by Lemma \ref{lemma:TiltedVarianceLowerBound}, is strictly positive. This completes the proof.
\end{proof}

\begin{lemma}[Asymptotic tilting regime] \label{lemma:LargeTiltingRegime}
    Let $Z$ be a stable source s.t. $\esup\limits_{P} Z = 0$. Then,
    \begin{align}
        \lim\limits_{\lambda \to \infty} E_{P_Z^{(\lambda)}}[Z] = 0.
    \end{align}
\end{lemma}
\begin{proof}
    We will first show that the tilted probability distribution concentrates near $0$ with increasing $\lambda$. For an arbitrary $\epsilon > 0$, consider
    \begin{align}
        P_{Z}^{(\lambda)}(Z \le -\epsilon) &= E_{P_{Z}^{(\lambda)}}\left[ \mathbf{1}(Z \le -\epsilon) \right]\\
        &= E_{P_{Z}}\left[ \exp\left( \lambda Z - \Lambda(\lambda)\right)\mathbf{1}(Z \le -\epsilon)\right]\\
        &\le \frac{\exp(-\lambda \epsilon)}{E_{P_{Z}}\left[ \exp(\lambda Z)\right]}\\
        &\le \frac{\exp(-\lambda \epsilon)}{E_{P_{Z}}\left[ \exp(\lambda Z)\mathbf{1}(Z \ge -\frac{\epsilon}{2})\right]}\\
        &\le \frac{\exp(-\frac{\lambda\epsilon}{2})}{P_{Z}(Z \ge -\frac{\epsilon}{2})}.
    \end{align}
    Therefore, for every $\epsilon$, we can find a sufficiently large $\lambda$ s.t. $P_{Z}^{(\lambda)}(Z \le -\epsilon)$ is arbitrarily small. Therefore,
    \begin{align}
        \lim\limits_{\lambda \to \infty} P_{Z}^{(\lambda)}(Z \le -\epsilon) = 0.
    \end{align}
    Now consider
    \begin{align}
        E_{P_Z^{(\lambda)}}\left[ Z \right] &= -E_{P_Z^{(\lambda)}}\left[ -Z \right]\\
        &= -\int\limits_{0}^\infty P_Z^{(\lambda)}(-Z \ge t) dt\\
        \implies \lim \inf\limits_{\lambda \to \infty}E_{P_Z^{(\lambda)}}\left[ Z \right] &= -\lim \inf\limits_{\lambda \to \infty} \int\limits_{0}^\infty P_Z^{(\lambda)}(-Z \ge t) dt\\
        &\ge -\lim \inf\limits_{\lambda \to \infty} \int\limits_{0}^\infty \frac{\exp(-\frac{\lambda t}{2})}{P_{Z}(Z \ge -\frac{t}{2})} dt.
    \end{align}
    Now, we can split the integral over the intervals $(0, \epsilon)$ and $(\epsilon, \infty)$:
    \begin{align}
        \lim \inf\limits_{\lambda \to \infty}E_{P_Z^{(\lambda)}}\left[ Z \right] &= -\lim \inf\limits_{\lambda \to \infty} \left( \int\limits_{0}^\epsilon \frac{\exp(-\frac{\lambda t}{2})}{P_{Z}(Z \ge -\frac{t}{2})} dt + \int\limits_{\epsilon}^\infty \frac{\exp(-\frac{\lambda t}{2})}{P_{Z}(Z \ge -\frac{t}{2})} dt \right)\\
        &\ge - \left(\int\limits_{0}^\epsilon \lim \inf\limits_{\lambda \to \infty}\frac{\exp(-\frac{\lambda t}{2})}{P_{Z}(Z \ge -\frac{t}{2})} dt + \lim \inf\limits_{\lambda \to \infty} \frac{1}{P_{Z}(Z \ge -\frac{\epsilon}{2})}  \int\limits_{\epsilon}^{\infty}\exp\left(-\frac{\lambda t}{2}\right) dt \right)\\
        &= 0.
    \end{align}
    In combination with the trivial upper bound $\lim \sup \limits_{\lambda \to \infty} E_{P_Z^{(\lambda)}}\left[ Z \right] = 0$, we obtain the required result.
\end{proof}

\subsection{Sum of independent sources}
Let $Z_1, Z_2,\cdots, Z_n$ be independent sources that satisfy (\ref{eq:stablesource1} - \ref{eq:stablesource2}). For any real $\lambda$ and $b$, we define
\begin{align}
    S_n &= \frac{1}{n}\sum\limits_{i=1}^nZ_i,\\
    \Lambda_i(\lambda) &= \Lambda_{Z_i}(\lambda),\\
    \Lambda(\lambda) &= \frac{1}{n}\sum\limits_i \Lambda_i(\lambda),\\
    \Lambda_n^*(b) &= \sup\limits_{\lambda} \left( \lambda b - \Lambda(\lambda) \right).
\end{align}
We will assume the existence of $c \in \mathbb{R}$ and a corresponding ${\eta^*}$ s.t. 
\begin{enumerate}[i)]
  \item $\Lambda(\lambda)<\infty$ in a neighborhood around ${\eta^*}$,
  \item ${\eta^*}>0$,
  \item $\Lambda'({\eta^*}) = c$, which is equivalent to stating that $\Lambda_n^*(c) = {\eta^*} c - \Lambda({\eta^*})$.
\end{enumerate}
We will also define the tilted variance $s_n^2 = \sum\limits_{i=1}^n \Var_{P_{Z_i}^{({\eta^*})}}\left[  Z_i \right]$ and the tilted third central moment $\mu^{(3)}_n = \sum\limits_{i=1}^n E_{P_{Z_i}^{({\eta^*})}}\left[ \Big| Z_i - E_{P_{Z_i}^{({\eta^*})}}[Z_i] \Big|^3 \right]$.
\subsection{Lemmas}

\begin{lemma} \label{lemma:ldMain}
    There exist uniform positive constants $\underline{M}({\eta^*}, \bar{Z}, \sigma^2)$ and $\overline{M}({\eta^*}, \bar{Z}, \sigma^2)$ s.t. for all sufficiently large $n$,
    \begin{align}
        \underline{M}({\eta^*}, \bar{Z}, \sigma^2) \le \frac{\Pr\left( S_n \ge c \right)}{\left(\frac{\exp\left(-n \Lambda_n^*(c)\right)}{\sqrt{n}}\right)} \le \overline{M}({\eta^*}, \bar{Z}, \sigma^2).
    \end{align}
\end{lemma}
\begin{proof}
    We start by proving the lower bound. Applying the result from \cite[Lemma~1]{altuug2020exact}, we obtain
    \begin{align}
        &\Pr\left( S_n \ge c \right) \nonumber\\
        &\ge \frac{\exp(-at_n)}{{\eta^*}\sqrt{2\pi}s_n}\left(1 - \frac{1}{a}\right)(1 + at_n)\left(1 - \frac{(1 + at_n)^2 + 1}{(1 + at_n){\eta^*}(1 - 1/a)2\sqrt{e}s_n}\right)\exp\left(-n \Lambda_n^*(c)\right),
    \end{align}
    where $a>1$ and $t_n = {\eta^*}2\sqrt{2\pi}\frac{\mu^{(3)}_n}{s_n^2}$. Choosing $a = 2$ and observing that $n\exp\left( -2{\eta^*} |\bar{Z}| \right)\sigma^2 \le s_n^2 \le n\bar{Z}^2$ and $\mu^{(3)}_n \le n\bar{Z}^3$, we directly obtain the required uniform lower bound for $n$ sufficiently large s.t. $\frac{(1 + at_n)^2 + 1}{(1 + at_n){\eta^*}(1 - 1/a)2\sqrt{e}s_n} \le \frac{1}{2}$.\\

    Next, for the upper bound, we have
    \begin{align}
        \Pr\left(S_n \ge c\right) &= E_{P_{Z^n}}\left[ \mathbf{1}\left\{ \frac{1}{n}\sum\limits_{i=1}^n Z_i \ge c\right\} \right]\\
        &= E_{P_{\tilde{Z}^n}}\left[ \exp\left( -{\eta^*} \cdot n \cdot S_n + n\Lambda({\eta^*}) \right) \mathbf{1}\left\{ \frac{1}{n}\sum\limits_{i=1}^n Z_i \ge c\right\} \right].
    \end{align}
    Substituting $W_i = -Z_i$, we obtain
    \begin{align}
        &\Pr\left(S_n \ge c\right) \\
        &= E_{P_{\tilde{Z}^n}}\left[ \exp\left( {\eta^*} \sum\limits_{i=1}^n W_i + n\Lambda({\eta^*}) \right) \mathbf{1}\left\{ \frac{1}{n}\sum\limits_{i=1}^n W_i \le -c\right\} \right]\\
        &= \exp\left(n\Lambda({\eta^*}) - cn\right)E_{P_{\tilde{Z}^n}}\left[ \exp\left( {\eta^*}  \sum\limits_{i=1}^n W_i + cn \right) \mathbf{1}\left\{ \frac{1}{n}\sum\limits_{i=1}^n W_i \le -c\right\} \right]\\
        &= \exp\left(-n\Lambda_n^*({\eta^*})\right)E_{P_{\tilde{Z}^n}}\left[ \exp\left( {\eta^*} \sum\limits_{i=1}^n W_i + cn \right) \mathbf{1}\left\{ \frac{1}{n}\sum\limits_{i=1}^n W_i \le -c\right\} \right]
    \end{align}
    We can then use Lemma 2 from \cite{altuug2020exact} to obtain 
    \begin{align}
        \Pr\left(S_n \ge c\right) &\le \left( \frac{1}{s_n\sqrt{2\pi}} + \frac{2\mu^{(3)}_n}{s_n^3} \right) \exp\left(-n\Lambda_n^*({\eta^*})\right).
    \end{align}
    Using the uniform bounds on $s_n$ and $\mu^{(3)}_n$ obtained previously, we can then derive $\overline{M}(\bar{Z}, \eta^*)$.
\end{proof}

\begin{lemma}[Refined Gibbs Conditioning Lemma] \label{lemma:GibbsBounded}
    There exists a uniform constant $C^*({\eta^*}, \bar{Z}, \sigma^2)$ s.t. for all sufficiently large $n$,
    \begin{align}
        E\left[ S_n | S_n \ge c\right] \le c + \frac{C^*({\eta^*}, \bar{Z}, \sigma^2)}{n}.
    \end{align}
\end{lemma}
\begin{proof}
    We have
    \begin{align}
        &E\left[ S_n | S_n \ge c\right]\\
        &=\frac{E_{P_{Z^n}}\left[ \frac{1}{n}\sum\limits_{i=1}^n Z_i \cdot \mathbf{1}\left\{ \frac{1}{n}\sum\limits_{i=1}^n Z_i \ge c \right\}\right]}{P_{Z^n}\left( \frac{1}{n}\sum\limits_{i=1}^n Z_i \ge c \right)}\\
        &= c + \frac{E_{P_{Z^n}}\left[ \frac{1}{n}\sum\limits_{i=1}^n (Z_i - c) \cdot \mathbf{1}\left\{ \frac{1}{n}\sum\limits_{i=1}^n (Z_i - c) \ge 0 \right\}\right]}{P_{Z^n}\left( \frac{1}{n}\sum\limits_{i=1}^n Z_i \ge c \right)}\\
        &\le c + \frac{E_{P_{Z^n}}\left[ \frac{1}{n}\sum\limits_{i=1}^n (Z_i - c) \cdot \mathbf{1}\left\{ \frac{1}{n}\sum\limits_{i=1}^n (Z_i - c) \ge 0 \right\}\right]}{\frac{\underline{M}({\eta^*}, \bar{Z}, \sigma^2)}{\sqrt{n}}\exp\left( -n\Lambda_n^*({\eta^*})\right)} \label{eq:lbLDProb}
    \end{align}
    Here, (\ref{eq:lbLDProb}) follows from the lower bound in Lemma \ref{lemma:ldMain}. Changing to the tilted measure, we then have
    \begin{align}
        &E\left[ S_n | S_n \ge c\right]\\
        &\le c + \frac{E_{P_{\tilde{Z}^n}}\left[\exp\left( -{\eta^*} \cdot n \cdot S_n + n\Lambda({\eta^*}) \right) \frac{1}{n}\sum\limits_{i=1}^n (Z_i - c) \cdot \mathbf{1}\left\{ \frac{1}{n}\sum\limits_{i=1}^n (Z_i - c) \ge 0 \right\}\right]}{\frac{\underline{M}({\eta^*}, \bar{Z}, \sigma^2)}{\sqrt{n}}\exp\left( -n\Lambda_n^*({\eta^*})\right)}\\
        &= c + \frac{E_{P_{\tilde{Z}^n}}\left[\exp\left( -{\eta^*} \cdot n \cdot S_n + {\eta^*} \cdot n \cdot c \right) \frac{1}{n}\sum\limits_{i=1}^n (Z_i - c) \cdot \mathbf{1}\left\{ \frac{1}{n}\sum\limits_{i=1}^n (Z_i - c) \ge 0 \right\}\right]}{\frac{\underline{M}({\eta^*}, \bar{Z}, \sigma^2)}{\sqrt{n}}}. \label{eq:NRDR}
    \end{align}
    Let us define $W_n = \frac{\sum\limits_{i=1}^n Z_i - c}{s_n}$ and $\psi_n = {\eta^*} s_n$. Following the same computations as in the proof of \cite[Lemma~V.2]{flamich2025redundancy}, we obtain
    \begin{align}
        &E\left[ S_n | S_n \ge c\right]\\
        &\le c + \frac{\sqrt{n}}{\underline{M}({\eta^*}, \bar{Z}, \sigma^2)\cdot {\eta^*}^2 \cdot n \cdot s_n}\\
        &\le c + \frac{1}{\underline{M}({\eta^*}, \bar{Z}, \sigma^2) \cdot {\eta^*}^2 \cdot \exp\left( -2{\eta^*} |\bar{Z}| \right)\sigma^2}\cdot\frac{1}{n}.
    \end{align}
\end{proof}

\end{document}